\documentclass[english,prl, longbibliography, reprint, twocolumn, groupedaddress]{revtex4-2}
\usepackage[T1]{fontenc}
\usepackage[latin9]{inputenc}
\usepackage{xcolor}
\usepackage{amsbsy}
\usepackage{amstext}
\usepackage{graphicx}
\usepackage{geometry}
\PassOptionsToPackage{normalem}{ulem}
\usepackage{ulem}

\makeatletter

\newcommand{\lyxmathsym}[1]{\ifmmode\begingroup\def\b@ld{bold}
  \text{\ifx\math@version\b@ld\bfseries\fi#1}\endgroup\else#1\fi}

\providecommand{\tabularnewline}{\\}

\providecolor{lyxadded}{rgb}{0,0,1}
\providecolor{lyxdeleted}{rgb}{1,0,0}
\DeclareRobustCommand{\mklyxadded}[1]{\textcolor{lyxadded}\bgroup#1\egroup}
\DeclareRobustCommand{\mklyxdeleted}[1]{\textcolor{lyxdeleted}\bgroup\mklyxsout{#1}\egroup}
\DeclareRobustCommand{\mklyxsout}[1]{\ifx\\#1\else\sout{#1}\fi}

\DeclareRobustCommand{\lyxdeleted}[4][]{\mklyxdeleted{#4}}

\usepackage{xcolor}
\definecolor{darkblue}{rgb}{0.1,0.2,0.6} 
\definecolor{lightblue}{rgb}{0.1,0.1,1.0}
\definecolor{darkred}{rgb}{0.8,0.1,0.2}
\usepackage[colorlinks,citecolor=lightblue,linkcolor=darkblue,urlcolor=lightblue] {hyperref}

\makeatother

\usepackage{babel}
\begin{document}
\title{Heavy-Tailed Hall Conductivity Fluctuations in Quantum Hall Transitions}
\author{Emuna Rimon, Eytan Grosfeld, Yevgeny Bar~Lev}
\affiliation{Department of Physics, Ben-Gurion University of the Negev, Beer-Sheva
8410501, Israel}
\date{December 30, 2025}
\begin{abstract}
We study the full distribution of the zero-temperature Hall conductivity
in a lattice model of the IQHE using the Kubo formula across disorder
realizations. Near the localization--delocalization transition, the
conductivity exhibits heavy-tailed fluctuations characterized by a
power-law decay with exponent $\alpha\approx2.3-2.5$, indicating
a finite mean but a divergent variance. The heavy tail persists across
a range of system sizes, correlation lengths of the disorder potential
and fillings. Our results demonstrate a breakdown of self-averaging
in transport in small, coherent samples near criticality, in agreement
with findings in random matrix models of topological indices.
\end{abstract}
\maketitle
\textit{Introduction}.--- Disorder plays a crucial role in the integer
quantum Hall effect (IQHE), giving finite widths to Hall conductivity
$\sigma_{xy}$ as a function of the filling or magnetic flux. While
scaling analysis has long established universal exponents for the
divergence of the localization length and the critical behavior of
transport \citep{Huckestein1995,MacKinnonKramer1981}, the statistical
properties of the transport observables themselves, beyond their disorder-averaged
values, remain much less explored. Understanding their full distribution
can reveal additional universal behavior and clarify whether, and
where, fluctuations become significant and may even dominate the transport.

In disordered metals and in the absence of a magnetic field, the longitudinal
conductance exhibits universal sample-to-sample fluctuations whose
variance is independent of microscopic details of the disorder \citep{altshuler1985fluctuations,LeeStone1985,lee1987universal}.
These universal conductance fluctuations (UCF) have a finite magnitude
of order $e^{2}/h$, implying that the conductance is not self-averaging
in one or two dimensions. Weak non-quantizing magnetic fields do not
significantly change the variance, however when magnetic field becomes
quantizing, specifically in two dimensions, the nature of conductance
fluctuations is less well understood. The Hall conductivity, $\sigma_{xy},$
is known to be self-averaging when it is quantized. Moreover, in gapped
or fully localized regimes the Hall conductivity also coincides with
the integer Chern number \citep{prodan2011disordered,kudo2019many}.
Large fluctuations therefore arise only at the plateau transitions,
where the chemical potential moves into a region of extended states.

Simulations of the Chalker-Coddington network model have shown that
the two-terminal conductance distribution at plateau transitions becomes
extremely broad, approaching a nearly uniform form. Such behavior
can indeed be observed in mesoscopic samples, where the sample dimensions
are not much larger than the dephasing length \citep{cobden1996measurement,cobden1999fluctuations}.
However, a two-terminal setup mixes longitudinal and transverse responses,
obscuring the intrinsic Hall physics. Even a nominal four-terminal
Hall measurement yields a value of $\sigma_{xy}$ inferred via the
macroscopic resistivity tensor, which in existing experiments averages
over many incoherent regions of the sample \citep{peled2003observation}.
Consequently, the full distribution of the microscopic Hall conductivity
- especially its tails, which control the convergence of its statistical
moments - remains essentially unexplored.

Spatially correlated disorder is known to smooth ensemble-averaged
$\sigma_{xy}$ profiles \citep{greshnov2007integer}, though it is
unclear whether this reflects suppressed fluctuations or simply narrower
transition widths. Recent analytical work on random matrices \citep{berry2018geometric,berry2020geometric,berry2020quantum},
predicts heavy-tailed distributions of geometric response functions,
suggesting that similar behavior may occur in disordered topological
phases. Numerical studies of disordered Majorana wires support this
perspective, revealing broad conductance distributions across disorder
realizations \citep{Antonenko2020}.

In this work, we compute the full distribution of the zero-temperature
Hall conductivity in a lattice model of the IQHE. We find that the
distribution develops heavy-tails near critical fillings, with a power-law
decay characterized by an exponent $\alpha\approx2.3\lyxmathsym{\textendash}2.5$,
implying a finite mean but a divergent variance. This behavior persists
across system sizes, disorder strengths, and disorder--correlation
lengths, demonstrating that it is an intrinsic and universal feature
of the critical regime rather than a finite-size artifact. These results
reveal a fundamental limitation of ensemble averaging near IQHE plateau
transitions and link the behavior of $\sigma_{xy}$ fluctuations to
recent predictions from random-matrix theory.

\begin{figure}
\includegraphics[width=1\columnwidth]{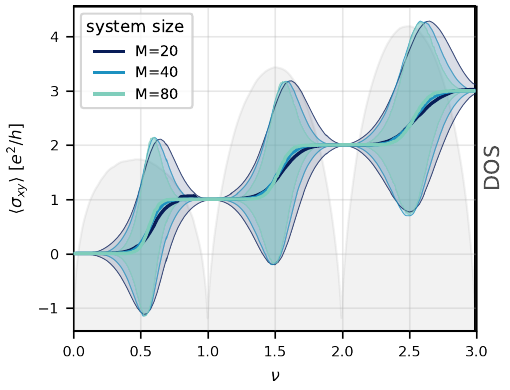}\caption{Ensemble mean and standard deviation of the Hall conductivity, $\sigma_{xy}(\nu)$,
with $W=1.0$ and $\eta=1.25$. Solid thick lines show the ensemble
mean, while solid thin lines denote the 95th percentile. The density
of states is shown in gray. Large realization-to-realization fluctuations
persist near transitions as $M$ increases.\label{fig:sigma_xy}}
\end{figure}

\textit{Model}.--- We study a tight-binding model of spinless electrons
on a two-dimensional square lattice subjected to a perpendicular magnetic
field, corresponding to a flux per plaquette of $\Phi_{\text{ }}/\Phi_{0}=1/10$
with $\Phi_{0}$ the flux quantum.

The Hamiltonian reads
\begin{equation}
H=-t\sum_{\langle\mathbf{r},\mathbf{r}'\rangle}e^{i\phi_{\mathbf{r}\mathbf{r}'}}\,c_{\mathbf{r}}^{\dagger}c_{\mathbf{r}'}+\text{h.c.}+\sum_{\mathbf{r}}V_{\mathbf{r}}\,c_{\mathbf{r}}^{\dagger}c_{\mathbf{r}},\label{eq:ham}
\end{equation}
where $t$ is the tunneling, $V_{\mathbf{r}}$ is the on-site random
potential, $c_{\boldsymbol{r}}^{\dagger}$ $\left(c_{\boldsymbol{r}}\right)$
creates (annihilates) a spin-polarized electron at site $\mathbf{r}=(x,y)$,
and $\phi_{\mathbf{r}\mathbf{r}'}$ is the Peierls phase, given in
the Landau gauge by $\phi_{\mathbf{r},\mathbf{r}+\hat{x}}=0$, $\phi_{\mathbf{r},\mathbf{r}+\hat{y}}=2\pi\,\frac{\Phi_{\text{}}}{\Phi_{0}}x$.
We take periodic boundary conditions on a square with $M\times M$
sites.

We consider two types of disorder: (i) uncorrelated white-noise disorder,
with $V\in\mathcal{U}[-W/2,W/2]$ (where $\mathcal{U}$ is the uniform
distribution), and (ii) Gaussian-correlated disorder characterized
by $\langle V(\mathbf{r})V(\mathbf{r}')\rangle=\frac{W^{2}}{12}\exp\left(-\frac{|\mathbf{r}-\mathbf{r}'|^{2}}{2\eta^{2}}\right)$
where $\eta$ is the disorder correlation length.

It is convenient to define the filling factor as $\nu=N/N_{\Phi}$,
where $N$ is the number of electrons and $N_{\Phi}=\Phi_{\text{tot}}/\Phi_{0}$
the total number of flux quanta through the system, which is also
the degeneracy of each Landau level in the continuum version of the
model \footnote{On the lattice Landau levels are only degenerate in the $\Phi/\Phi_{0}\to0$
limit.}. When the system is clean, the Hall conductivity is quantized whenever
an integer number of Landau levels is completely filled \citep{thouless1982quantized}.\lyxdeleted{Eytan Grosfeld}{Mon Dec 29 10:16:55 2025}{
} For most partial fillings the Hall conductivity remains quantized,
since the addition of disorder localizes all states, except a critical
state at the center of each Landau band, and the localized states
do not contribute to transport \citep{chalker1987anderson}.

The dimensionless zero-temperature Hall conductivity is computed using
the Kubo formula \citep{Kubo1957,Huckestein1995}:
\begin{equation}
\sigma_{xy}=\frac{2\pi i}{M^{2}}\sum_{n\neq0}\frac{J_{y}^{0n}J_{x}^{n0}-J_{x}^{0n}J_{y}^{n0}}{\left(E_{n}-E_{0}\right)^{2}},\label{eq:kubo}
\end{equation}
where $J_{a}^{0n}=\langle0|J_{a}|n\rangle$ is the matrix element
of the $a=x,y$ component of the current operator $\left(J_{a}\right){}_{\boldsymbol{r},\boldsymbol{r}+\hat{a}}=\partial H/\partial\phi_{\boldsymbol{r},\boldsymbol{r}+\hat{a}}$
between the ground state $|0\rangle$ with energy $E_{0}$ and the
excited state $|n\rangle$ with energy $E_{n}$, and $J_{a}^{n0}=\left(J_{a}^{0n}\right)^{*}$.
This expression is formally derived from linear-response theory, and
its interpretation as the physical Hall conductivity strictly requires
an energy gap separating the occupied and unoccupied states \citep{akkermans1997twisted}.
The existence of a gap ensures adiabatic response and allows the DC
limit to be taken smoothly. In regions where the spectrum becomes
gapless, its structure continues to encode the Berry curvature of
the occupied states, but it no longer corresponds to the conductivity.
In such cases, $\sigma_{xy}$ as extracted from the Kubo formula should
be viewed as a measure of local geometric response, rather than a
transport coefficient \citep{de2024derivation}.

\emph{Results}.--- Fig. \ref{fig:sigma_xy} shows the mean $\sigma_{xy}(\nu)$,
computed using Eq. (\ref{eq:kubo}) for about 100,000 disorder realization
and the 95th percentile of its distribution. We recognize significant
sample-to-sample fluctuations that persist as the linear system size
$M$ increases from 20 to 80. This lack of self-averaging is most
prominent near the transition regions (e.g. $\nu=2.5$), where the
density of states is the largest, while the plateau regions (e.g.
$\nu=2.0$), where there are few states, appear comparatively stable.

The lack of convergence with increasing system size $M$ at half fillings,
suggests that increasing the system size is insufficient to recover
a typical $\sigma_{xy}$ from one disorder sample at the critical
regimes. The width of the critical regime, where the distribution
of conductivity is broad, does appear to decrease with system size
\citep{author2025supplement}.

\begin{figure}
\includegraphics[width=1\columnwidth]{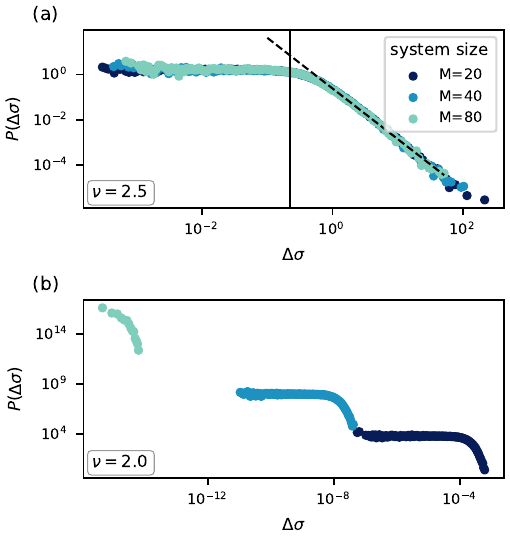}\caption{Distribution of Hall conductivity deviations, $P\!\left(|\sigma_{xy}-\langle\sigma_{xy}\rangle|\right)$,
presented in log--log scale, at $\nu=2.5$ (a) and $\nu=2.0$ (b)
with $\eta=0.75$. The dashed line is the fitted power law slope and
the vertical solid line indicates where the tail begins.\label{fig:pdf}}
\end{figure}

To quantify the fluctuations, we examine the full distribution of
conductivity deviations across disorder realizations, $P(\Delta\sigma)$,
where $\Delta\sigma=|\sigma_{xy}-\langle\sigma_{xy}\rangle|$, and
$\left\langle \sigma_{xy}\right\rangle $ is the mean. Near the critical
filling $\nu=2.5$, the distribution exhibits a heavy tail consistent
with a power-law decay, Fig. \ref{fig:pdf}a. This behavior persists
across system sizes, indicating that the fluctuations are not suppressed
in the thermodynamic limit. We characterize the distribution by the
power-law exponent of its tail $\alpha$, as well as the point at
which the power-law tail emerges, $\Delta\sigma_{0}$. In contrast,
on the plateau ($\nu=2.0$, for example), the distribution is extremely
narrow, and becomes increasingly so with system size, Fig. \ref{fig:pdf}b.

\begin{figure}
\includegraphics[width=1\columnwidth]{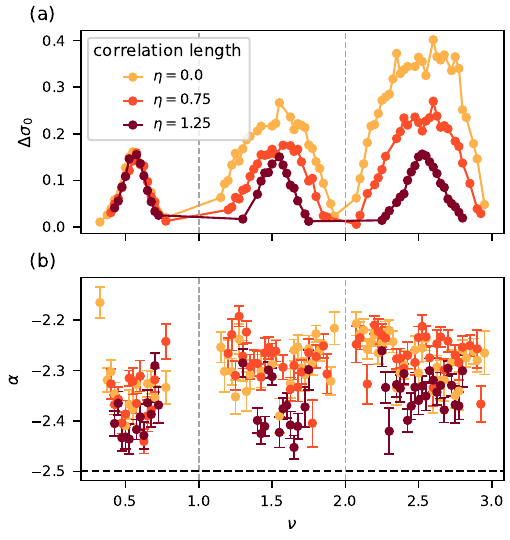}\caption{Fitted distribution width $\Delta\sigma_{0}$ (a) and power-law exponent
$\alpha$ (b) from a distribution function of Hall conductivity. The
dashed horizontal line is the power law $-2.5$ predicted by Berry\citep{berry2018geometric,berry2020geometric,berry2020quantum}
using random matrix models of topological indices. \label{fig:effect-of-correlations}}
\end{figure}

To characterize the effect of disorder correlations, we present in
Fig. \ref{fig:effect-of-correlations} the fitted exponent $\alpha$
and offset $\Delta\sigma_{0}$ as a function of the filling $\nu$
and the disorder correlation length, $\eta$ for a fixed $M=80$.
By characterizing each distribution with these two parameters, we
identify the fillings for which $\alpha$ and $\Delta\sigma_{0}$
remain stable with increasing system size. Remarkably, we find that
while spatial correlations reduce $\Delta\sigma_{0}$ and make the
distribution more narrow, they do \emph{not} regularize the tail.
The power-law exponent $\alpha$ remains nearly constant across the
critical region, possibly displaying a non-significant reduction with
$\eta$ for some filling factors. This indicates that the breakdown
of self-averaging persists in the presence of correlated disorder,
but may explain previous numerical results showing the apparent smoothing
of plateau transitions, hinged in finite-size effects \citep{greshnov2007integer}.
In the vicinity of integer fillings the distribution no longer exhibits
a flat central region, and the power-law description fails. As a result,
our fitting procedure does not yield meaningful $\alpha$ or $\Delta\sigma_{0}$
values, and no points appear in this part of the figure.

The persistence of power-law tails with $\alpha<3$ implies that while
the mean $\langle\sigma_{xy}\rangle$ is finite the variance diverges.
The breakdown of self-averaging of Hall conductivity fluctuations
near the critical point is our main result.

\textit{Discussion}.--- We studied the distributions of zero temperature
Hall conductivity across the Hall transitions by varying the filling
of the system, the correlation of the disorder and the system size.
The distributions we obtain display power-law tails with exponents
$\alpha\approx2.3\lyxmathsym{\textendash}2.5$, implying divergent
variance and scale-invariant behavior.

The exponent of the power-law is persistent across system sizes, disorder
strengths, and correlation lengths indicating that it originates from
the critical regime of the integer quantum Hall transition rather
than from finite-size effects. For fillings where the Hall conductivity
is quantized it is known to correspond to a topological invariant
\citep{kudo2019many}. While this correspondence breaks at the transition,
the zero temperature Kubo formula for Hall conductivity, still represents
the local curvature of the filled single-particle states \citep{xiao2010berry},
linking the statistical fluctuations in the conductivity to the topological
character of the transition. Correlations do reduce the width of the
distributions and narrow the transition region \citep{greshnov2007integer},
but our results demonstrate that they do not regularize the heavy
tail itself.

In contrast to the universal conductance fluctuations, which exhibit
a finite, size-independent variance in diffusive metals, and has been
verified in mesoscopic transport experiments, the Hall conductivity
in the IQHE shows non-universal, heavy-tailed behavior near criticality.
This highlights the difference between transport in regular and topological
systems where heavy-tailed distributions reflect rare, topology-changing
disorder configurations. Notably, macroscopic Hall-bar experiments
starting from the work of von Klitzing \citep{klitzing1980new,Huckestein1995}
show smooth transitions between Hall-resistance plateaus, without
the fluctuations seen in mesoscopic devices. This suggests that our
model corresponds more closely to the mesoscopic regime, where fluctuations
do appear in the transition region \citep{cobden1996measurement,cobden1999fluctuations}.
Even there the heavy tail and its universal exponent is not a characteristic
of the two-terminal conductance, but instead should be accessed via
a four-terminal measurement in small, coherent samples.

The emergence of heavy-tailed statistics in our results is reminiscent
of the distributions found by Berry and Shukla in random matrix ensembles
\citep{berry2018geometric,berry2020geometric,berry2020quantum}, although
the contexts differ. In their work, power-law behavior arises from
generic features of random quantum geometry, whereas in our case it
appears in a concrete transport setting, within a physical topological
model. The similarity suggests that heavy-tailed statistics may be
a broader feature of complex quantum systems.

In summary our results demonstrate that self-averaging breaks down
for a topological observable $\sigma_{xy}$ in the critical regime
and suggest that heavy-tailed statistics may be a generic feature
of disorder-driven topological phase transitions, extending beyond
conventional transport phenomena.

\bibliography{bib}

\clearpage{}

\pagebreak{}

\clearpage
\setcounter{section}{0}
\setcounter{equation}{0}
\setcounter{figure}{0}
\setcounter{table}{0}
\renewcommand{\thesection}{S\arabic{section}}
\renewcommand{\theequation}{S\arabic{equation}}
\renewcommand{\thefigure}{S\arabic{figure}}
\renewcommand{\thetable}{S\arabic{table}}
\begin{widetext}

\begin{center}
    \Large \textbf{Supplementary Material for ``Heavy-Tailed Hall Conductivity Fluctuations in Quantum Hall Transitions''}\\[1em]
    \normalsize Emuna Rimon, Eytan Grosfeld, Yevgeny Bar Lev\\[0.5em]
    \emph{Department of Physics, Ben-Gurion University of the Negev, Beer-Sheva, 84105, Israel}\\[1em]
\end{center}

\end{widetext}

\section{Width change}

In Fig. 1 we observed that the width of the transition region, as
a function of filling factor, becomes narrower with increasing system
size.

For each of the three lowest Landau bands, we quantify how this transition
width depends on system size and on the disorder correlation length.

\begin{figure}[h]
\includegraphics[width=1\columnwidth]{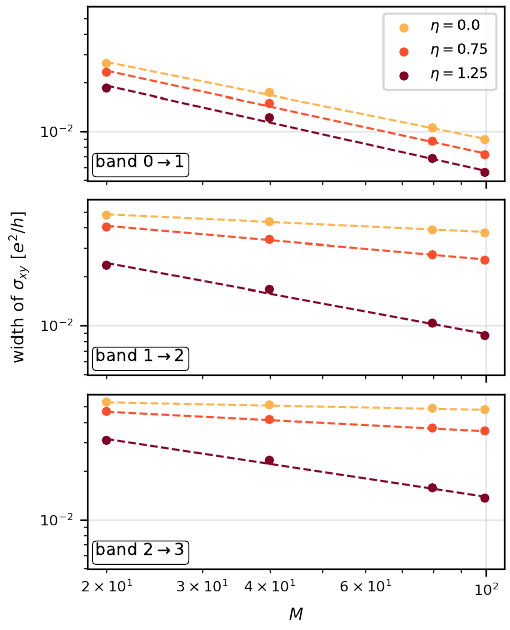}\caption{The widths of distributions defined as he $95$th percentile of the
Hall conductivity $\sigma_{xy}(\nu)$. In each figure, we show the
second moment of these widths between consecutive integer fillings.
\label{fig:widths}}
\end{figure}

As seen in Fig. \ref{fig:widths}, the transition width decreases
approximately as a power law with system size. Introducing spatial
correlations slightly enhances this trend, leading to a faster decay
of the width with system size.

\begin{table}[h]
\bigskip{}
\begin{tabular}{|c|c|c|c|}
\hline 
 & $\eta=0.0$ & $\eta=0.75$ & $\eta=1.25$\tabularnewline
\hline 
\hline 
band $0$ & $0.680\pm0.025$ & $0.733\pm0.028$ & $0.753\pm0.043$\tabularnewline
\hline 
band $1$ & $0.157\pm0.007$ & $0.295\pm0.012$ & $0.626\pm0.626$\tabularnewline
\hline 
band $2$ & $0.066\pm0.003$ & $0.173\pm0.002$ & $0.514\pm0.032$\tabularnewline
\hline 
\end{tabular}\caption{Values of slopes $-\alpha$ for the decay of the second moment of
these widths between consecutive integer fillings with system size.}
\end{table}

\end{document}